# Three Mode Interactions as a Precision Monitoring Tool for Advanced Laser Interferometers

**L Ju, C Zhao, D G Blair, S Susmithan, Q Fang and C D Blair**
School of Physics, The University of Western Australia
35 Stirling Highway, Crawley, WA 6009, Australia

Email: li.ju@uwa.edu.au

**Abstract**. Three-mode opto-acoustic interactions in advanced laser interferometer gravitational wave detectors have high sensitivity to thermally excited ultrasonic modes in their test masses. Three mode interaction signal gain can change by 100% for thermally induced radius of curvature variations $\sim 10^{-5}$, allowing the monitoring of thermal distortions corresponding to wavefront changes $\sim 2 \times 10^{-13}$m. We show that the three-mode gain for single cavity interactions can be monitored by observing beat signals in the transmitted or reflected light due to the thermal excitation of the many hundreds of detectable acoustic modes. We show that three mode interaction signals can be used at low optical power to predict parametric instabilities that could occur at higher power. In addition, at any power, the observed mode amplitudes can be used to control the interferometer operating point against slow environmental perturbations. We summarize data on an 80m cavity that demonstrates these effects and propose testing on full scale interferometer cavities to evaluate whether the technique has practical benefits that can be extended from single cavities to dual recycling interferometers.

## 1. Introduction

The first generation of large scale laser interferometer gravitational wave detectors reached their design sensitivity of $h \sim 3 \times 10^{-23}/\sqrt{Hz}$, sufficient to detect rare astronomical events such as black holes mergers and neutron star coalescence events within 10-100Mpc [1]. Advanced detectors being constructed are aiming for sensitivity $h \sim$ few $\times 10^{-24}/\sqrt{Hz}$, sufficient to observe the estimated population of neutron star coalescence events at a rate of tens of events per year.

This sensitivity improvement requires the use of high laser power inside the optical arm cavities to reduce shot noise, possibly in combination with optical squeezing [2]. High optical power leads to enhanced three mode opto-acoustic interactions between the main cavity $TEM_{00}$ mode, cavity transverse modes and acoustic modes of the suspended test mass mirrors. Rare interactions for which the opto-acoustic overlap factor is large [3,4] can achieve parametric gain R>1, which leads to parametric instability. Precise quantitative estimates are difficult because of uncertainties in material parameters, mirror parameters, optical alignments and losses [5]. Modelling to date indicate that a few acoustic modes per test mass could become unstable [6].

As discussed below, modelling also indicates that ~1000 acoustic modes in the frequency range 10kHz to 150kHz, interacting with ~100 optical modes have parametric gain $>10^{-3}$ [6]. This is sufficient that the thermally excited acoustic modes can be easily observed by the beating signal in the transmitted or reflected light. Such signals are detectable using a spatially sensitive photodetector such as a quadrant photodetector (QPD). Because the cavity transverse mode frequencies are very sensitive to mirror radius of curvature

(RoC), which itself is tuned by absorbed laser light, and because the optical line widths are very narrow, the parametric gain (and hence the mode amplitude observed in the beating signal) is very sensitive to the interferometer operating conditions [7].

Three mode parametric interactions have been extensively studied [8] in an 80m optical cavity at the Gingin high optical power facility [9]. By thermally changing the radius of curvature of a cavity test mass mirror, using either a thermal compensation plate [10] or a $CO_2$ laser [11], parametric interaction signals have been observed with high signal to noise ratio. Three mode parametric instability in a free space optical cavity was first reported by Chen *et al*. [12], using a very low mass resonator, but has still not been observed in a long baseline interferometer cavity.

Many techniques for controlling parametric instability have been considered and investigated [13,14,15,16,17]. To some extent this work is hampered by lack of knowledge of which acoustic and optical modes are likely to be unstable, since this depends very strongly on unpredictable details such as thermal deformations and alignment variations. It would be very useful to have a means of diagnosing and predicting parametric instability before it occurs.

The monitoring of three mode interactions (3MI monitoring) for the purpose of parametric instability suppression has already been proposed and demonstrated for a single mode. This is the method of optical feedback control [14,18]. It requires the monitoring of an acoustically excited high order optical mode, externally generating an out of phase optical signal, and re-injecting it to the cavity, so that interference suppresses the offending mode.

In this paper we address a simpler, but much more general approach involving 3MI monitoring of numerous mostly low gain acoustic modes that are detected through three mode interactions. We propose two ways in which this monitoring can be useful. Firstly it can be used to obtain advance warning of parametric instabilities that are likely to occur at higher optical power. This can allow detector commissioners time to design specific control strategies for specific predicted instabilities. Secondly, the method can provide error signals which might be useful in controlling alignments and inhomogeneous temperature gradients that may not otherwise be easily measured.

The proposed method relies on the fact that all acoustic modes are thermally excited and very well vibration isolated. For this reason the kT mean thermal energy of each mode can be treated as a calibration signal. As long as the mode energy is integrated over several relaxation times (typically integrating for 10-$10^3$ seconds) it provides a calibration signal with precision that increases with integration time. If the parametric gain approaches unity the calibration must be corrected by the parametric mode amplification which acts to increase the mode temperature by a factor 1/1-R.

The sensitivity of 3MI monitoring arises from the sensitivity of parametric interaction gain to thermally induced test mass radius of curvature changes, and test mass alignment. For example, in a single arm cavity, the gain of some interactions changes by up to ~ 100% for a few centimeters change in the ~2000m RoC. The latter change may be created by very small changes in mirror heat load.

Advanced LIGO and Virgo already plan an extensive program of wavefront monitoring using Hartmann sensors to detect thermal distortions of test masses. The 3MI monitoring proposed here does not replace the use of Hartmann sensors, nor other sensors such as beam spot imaging and optical levers. At minimum it can provide early experimental prediction of parametric instability, but it has the possibility of providing further useful information.

There are several factors that give 3MI monitoring an ability to diagnose multiple degrees of freedom in an interferometer including radii of curvature, inhomogeneous thermal distortion and optical alignments.
a) Each acoustic mode signal is the result of an opto-acoustic overlap between an acoustic mode and one or more high order transverse cavity modes. The overlap depends strongly on the high order mode position relative to the test mass, so transverse mode positions and orientations can be estimated.
b) Because parametric gain depends linearly on input laser power, monitoring of 3MI gain at low power can allow prediction of parametric instability at a higher power.
c) The high order transverse modes sample larger and different areas of the test mass compared with the main $TEM_{00}$ beam. Hence 3MI monitoring can in principle detect thermally induced inhomogeneous radius of curvature variations.
d) Each acoustic mode signal is associated with an individual test mass. The large acoustic mode frequency dependence on temperature, $df/f \sim 10^{-4} K^{-1}$ enables individual modes to be identified simply by observing the tuning when the thermal environment is altered (eg by warming the vacuum envelope of the test mass suspension tank). In principle the thermal state of a test mass could be monitored or controlled by comparing the 3MI gains of several acoustic modes.

e) As we show below, ~700 acoustic modes should be able to be monitored simultaneously. These signals are not all independent, because many have common associated transverse modes. However since many acoustic mode shapes can be inferred based on finite element modeling, unique solutions for important interferometer control parameters may be obtainable.

f) Parametric gains also depend on other interferometer components such as the beam splitter, the power and signal recycling mirrors, and power recycling cavity compensation plates, which all affect the transverse mode resonance. These components represent additional degrees of freedom that might be able to be measured through multiple 3MI monitoring.

Thus we suggest that a program of 3MI monitoring could be valuable, especially because the enormous thermal memory of fused silica test masses (due to their long thermal relaxation time) means that they will almost never be in a state of dynamic equilibrium. However the usefulness of 3MI monitoring will depend on the signal to noise ratio achievable, the stability of measurements, and quality of finite element models, and the ability to find unique solutions.

In this paper we will show that 3MI monitoring is clearly a useful tool for single cavities. We present modelling data that shows that it can in principle be extended to a full dual recycling interferometer. However none of the modeling takes into account the asymmetric imperfections in the 8 independent optical elements mentioned in point f) above that play a role in each three mode interaction. Thus we are unable to prove that the method proposed will be a practical tool in operating advanced interferometers.

We report on a combination of advanced interferometer modeling and experiments on 80m optical cavities designed to achieve conditions similar to advanced interferometers. We will show that 3MI monitoring can identify modes that will become unstable at higher optical power in single cavities, and show that 3MI monitoring is very sensitive to thermal conditions and to beam spot position. Finally we will discuss a readout scheme that could be implemented on advanced interferometers and suggest avenues for further investigation. First we will review the theory that underpins this work.

## 2. Theory of three mode interactions and predictions for Advanced detectors

In an optical cavity, the thermal motion of a test mass internal mode scatters the cavity $TEM_{00}$ mode into higher order transverse modes $TEM_{mn}$ (m and n are the mode number). The optical modes in turn interact with the test mass via the radiation pressure produced by the beating of the carrier and the transverse mode. Because this interaction involves 2 optical modes—the $TEM_{00}$ and $TEM_{mn}$ modes and one acoustic mode, it is called a three mode parametric interaction. The process can lead to either parametric cooling of the acoustic mode or parametric amplification (heating) of the acoustic mode. The gain of the 3-mode parametric interaction is given by [3]

$$R = \frac{I}{M\omega_0\omega_m^2 L^2} \frac{\Lambda}{1+(\Delta\omega_s/\delta)^2} Q_0 Q_1 Q_m . \quad (1)$$

Here $I$ is the laser power, $Q_0$, $Q_1$, and $Q_m$ are the quality factors of the two optical modes and the acoustic mode of the mechanical resonator, m is the effective mass of the test mass acoustic mode, $\omega_0$ and $\omega_m$ are the frequencies of the cavity and the resonator respectively, $L$ is the length of the cavity, $\delta$ is the half width of the high order mode. The factor $\Lambda$ is the mode shape overlap integral [3], describing how well the acoustic mode shape matches the optical mode shape. The factor $\Delta\omega_s = \Delta\omega - \omega_m$, where $\Delta\omega$ is the frequency difference between the $TEM_{00}$ mode, $\omega_0$, and the high order mode $\omega_a$, is given by

$$\Delta\omega = \omega_0 - \omega_a = \frac{c}{L}(m+n)\arccos\left(\pm\sqrt{\left(1-\frac{L}{R_1}\right)\left(1-\frac{L}{R_2}\right)}\right). \quad (2)$$

Here $R_1$ and $R_2$ are the radii of curvature of the end mirrors of the cavity. The ± sign depends on the cavity configuration.

Equation (1) is simplified, considering only the Stokes process where parametric amplification or instability processes occur with only one higher order optical mode taken into account. There are a roughly equal number of negative gain modes that can be equally useful for obtaining interferometer information but which will not cause instability. Furthermore in many cases multiple higher modes can interact with a single acoustic mode, so that the gain *R* is a summation over several 3MI processes [19].

It can be seen from equation (1) that in a high optical power cavity with high optical Q factors, when the frequency condition $\omega_0 - \omega_a = \omega_m$ is met and the high order mode has high spatial overlap with the acoustic mode, the parametric gain can be large, leading to parametric instabilities where the acoustic mode amplitude will ring up exponentially. The overlap parameter is a key parameter here because it is sensitive to the position of the optical cavity transverse mode relative to the acoustic mode which is fixed in the test mass. Small motion of the beam position or distortion of the transverse mode can change the parametric gain strongly.

Gras *et al* [6] analysed parametric instabilities in advanced gravitational wave detectors and showed that there would always be a few unstable modes in each test mass at full operating power unless the parametric instability is suppressed. Unfortunately 3MI is so sensitive to system parameters such as laser spot position, mirror radius of curvature, and actual mode shapes (which can be significantly distorted by mirror thermal distortions) that the specific modes that will become unstable are almost impossible to predict in advance. However it is also this sensitivity to optical parameters that enables 3MI monitoring to be a useful tool.

## 3. Modelling Studies

Figure 1(a) shows a snapshot of acoustic modes of a test mass in a 4km cavity with a particular radius of curvature (which defines the transverse mode spectrum). The figure shows the 3MI gain as a function of frequency. Higher gain modes tend to cluster around the transverse mode offset frequencies. Those points with R>1 would normally be supressed to below unity for stable operation of the interferometer. If the radius of curvature changes or if the laser spot moves on the mirror, the points representing the parametric gain move up or down such as the exampes discussed below. Typically, for a fused silica test mass for advanced detectors, we expect about 700 modes per test mass for R> $10^{-3}$ as shown in Figure 1(b). We choose R~$10^{-3}$ as a threshold because this corresponds to easy detectability in experimental studies.

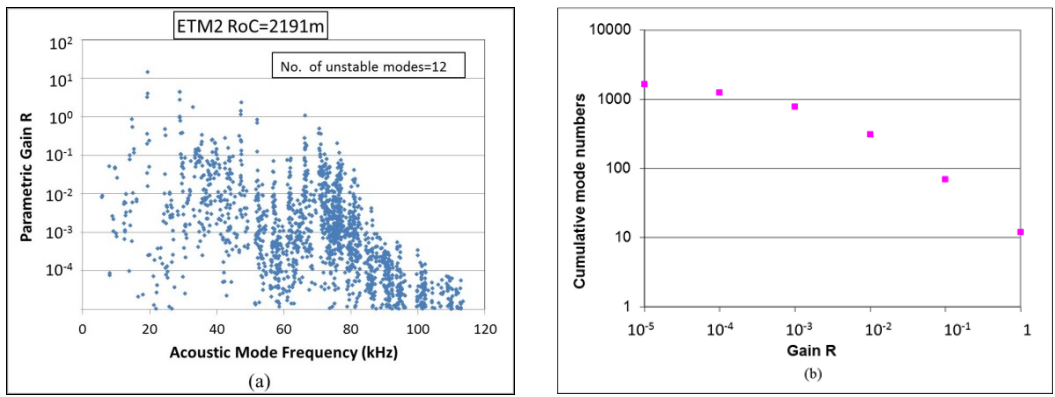

**Figure 1**. (a) An example of an acoustic mode showing the variability of parametric gain in a 4km advanced gravitational wave detector configuration with RoC=2191m [20]. (b) The cumulative number of acoustic modes with gain above a given value in an advanced LIGO type interferometer with fused silica test masses. All modes with gain > $10^{-3}$ should be easily detectable with a multi-element photodetector.

Figure 2 shows a typical model results for 3MI gain as a function of RoC for an advanced interferometer configuration. It can be seen that some modes change their gain by several orders of magnitude over a few meters change in radius of curvature and some show asymptotic behavior in which the gain can change by 100% in a few cm change in RoC. Amongst the large density of modes there are always a few with such high sensitivity to radius of curvature

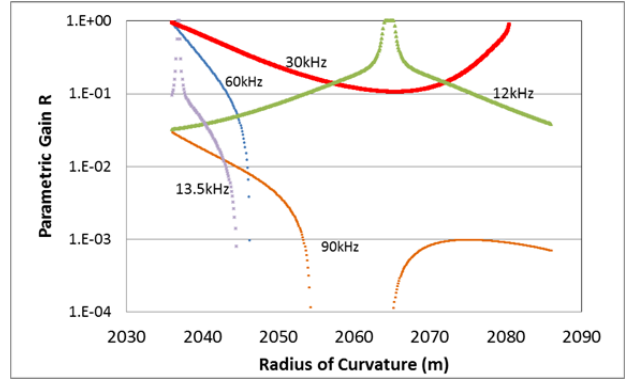

**Figure 2**. The parametric gain R as a function of mirror radius of curvature for several representative modes based on the same modelling as Fig.1.

It is useful to calculate the magnitude of the wavefront distortions corresponding to the observed radius of curvature sensitivity discussed above. A change in radius of curvature translates to a fractional change in wavefront according to the relation given in equation (3) below.

$$\Delta d = \frac{r^2}{2R_m^2} \Delta R_m, \qquad (3)$$

where $R_m$ is the mirror radius of curvature, and $r$ is the effective radius of the test mass and $d$ is the depth of the mirror deformation. For $\Delta R_m$ ~2mm, $\Delta d$ ~2 × $10^{-13}$m and for $\lambda$ =1064nm radiation, this means that 3MI has the capability of monitoring wavefront changes ~ 2 × $10^{-7}$ $\lambda$. In this sense 3MI offers monitoring at an unprecedented precision.

## 4. Three mode interaction monitoring at Gingin

We have investigated 3MI in two different ~80m cavities at the Gingin High Optical Power Facility (HOPF). The systems are rather simple. A cavity with high g-factor is locked using about 2W of $TEM_{00}$ injected 1064nm light. A quadrant photodetector monitors the transmitted beam. A $CO_2$ laser is used to apply variable heating to the centre of the end test mass, thereby creating a temperature gradient that deforms the mirror. The south arm of the Gingin facility uses a cavity with sapphire test masses while the east arm uses fused silica test masses. Both cavities have allowed observation of three mode interactions.

Because the test masses in both cavities are relatively small, the acoustic mode density is low and instead of seeing a large number of modes simultaneously we observe modes only when the $CO_2$ laser power has correctly tuned the transverse mode frequency to allow a particular mode to be observed.

Modes with parametric gain ~$10^{-3}$ are relatively easily observed. Normally monitoring is done using a quadrant photodetector which is particularly sensitive to the $TEM_{01}$ mode. For the monitoring of a more general set of modes, it would be advantageous to use a 9-element photodetector with ability to sum and difference channels so as to have maximum sensitivity to a particular optical mode shape.

In the south arm sapphire test mass cavity, Zhao, *et al* first observed a 3MI interaction with a gain ~ 0.01 when the radius of curvature of the end test mass was appropriately thermally tuned [21]. The thermal peak has an amplitude of a few parts in $10^{-15}$ m (corresponding to kT of energy), and the noise floor is about $10^{-17} m/\sqrt{Hz}$. From this observation we can predict from equation (1) that if the input laser power was increased from 2W to 200W the mode would reach the instability threshold, with $R$~1.

By sweeping through the RoC of a sapphire test mass in the south arm cavity, using $CO_2$ laser heating [22], acoustic mode thermal peaks appear with the change of the heating power, as shown in Figure 3. A few hundred milliwatts change in heating power is sufficient to sweep across the entire resonant peak. Typically the 3MI gain changes by a factor 3 as the heating power changes by ~100mW.

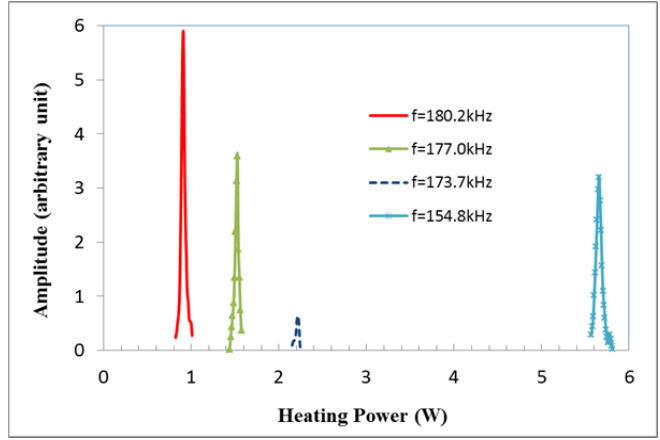

**Figure 3.** Three mode amplification of thermal noise of test mass acoustic modes during controlled thermal tuning using a $CO_2$ laser. Here the $CO_2$ laser is tuning the transverse mode frequency and thereby changing the three mode tuning condition. Peak height variations correspond to different overlap factors.

Another observation was that the position and size of the $CO_2$ laser spot significantly alters the 3MI signal. In this case we assume that we are observing inhomogeneous changes in the shape of the mirror which was distorting both the transverse mode shape and the mode frequency.

The east arm system was designed specifically to study 3MI. It is a near concentric cavity similar to an Advanced LIGO arm cavity, with fused silica test masses 100mm diameter × 50mm thickness in a cavity of length 73.9m, and radii of curvature of 37.4 m and 37.5 m. The measured cavity finesse is 14500+/-300, and the cavity g-factor is ~0.98, leading to a transverse mode spacing from Eq. (2), of ~100kHz. This means that test mass acoustic modes at ~100 kHz can scatter the cavity fundamental mode into the cavity first order mode while acoustic modes near 200 kHz will scatter the fundamental mode into the second order mode, etc. Thermal tuning using $CO_2$ laser heating can decrease the g-factor to 0.95, increasing the mode spacing to about ~ 150 kHz.

Without $CO_2$ heating and at a cavity power level of ~ 3kW, we observed that a few millimeters change in position of the laser spot on a test mass causes easily detectable changes in the 3MI signal, which is due to changes in the overlap parameter. This demonstrates the high sensitivity of the 3MI signal to laser beam alignments. This is expected because misalignment changes the position of optical anti-nodes relative to acoustic anti-nodes. Changes in overlap can cause large reductions in the overlap parameter such that some modes fall below the noise floor while different acoustic modes become visible. Another cause of the 3MI signal change could be due to the fact that mirrors in east arm are not of the highest quality so that beam spot misalignment would also incur cavity g-factors change due to the imperfection of the radii of curvature of the mirrors and inhomogeneous optical absorption. Thus it is possible to use relative mode amplitudes to diagnose changes of the optical cavity conditions such as g-factor, and spot position.

Since the acoustic modes have moderately high Q-factor, the acoustic mode frequency is easily measured. As we increase the cavity power from 500W to 5kW in the fused silica east arm cavity, we observed three mode interactions similar to the $CO_2$ tuning result in the south arm described above. The carrier laser heating causes self-induced thermal lensing, thereby shifting the $TEM_{01}$ mode frequency which determines the frequency at which acoustic modes are resonant. At low power we can observe a 94kHz acoustic mode, but as the cavity power increases this ceases to be detectable, while other modes appear at ~105kHz and then 112kHz, as shown in Figure 4. By using a $CO_2$ laser and correlating frequency with $CO_2$ laser heating, this provides an easy means of identifying the test mass associated which a particular acoustic mode.

We have also observed strong thermal tuning of the acoustic modes in our fused silica test masses due to the temperature dependence of Young's Modulus. The frequency tuning coefficient $\delta f/f$ in fused silica is anomalous and positive, and has a magnitude $\sim 10^{-4}$ per degree. The acoustic mode frequencies increase with time due to the increasing average temperature of the test mass.

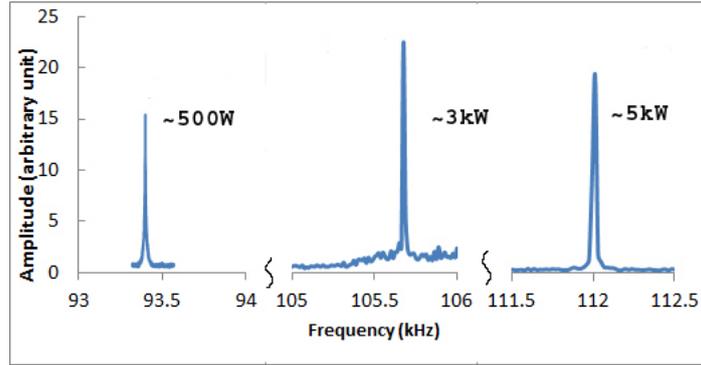

**Figure 4**. Three acoustic modes of the fused silica test masses with Q-factor ~$4\times10^4$, observed by three mode interaction. In this case the cavity transverse mode is tuned by optical absorption in the test masses, causing different acoustic modes to be resonant.

5. **Conclusions**

We have shown that the three mode interactions normally present in large scale gravitational wave detector arm cavities can be used to create a precision monitoring tool which is sensitive to small variations in mirror radii of curvature and spot positions. The high sensitivity of acoustic mode readout to small changes in test mass absorbed power has been demonstrated by both modelling and observations in two optical cavities. The method could be useful for providing ancillary data to enable stable operation of high optical power interferometers in the presence of slow thermal fluctuations.

Experimental results have been presented that demonstrate the ability for low power measurements to predict modes likely to become unstable at high optical power in a single arm cavity. Both alignment and thermal perturbations in a single cavity are easily observed. In a more complex advanced interferometer the parametric gain depends on the Guoy phase associated with the entire coupled cavity configuration. The idealised modelling presented (that neglects the thermal compensation masses and optical inhomogeneities) shows that the method carries over to full dual recycling interferometers, but the large number of degrees of freedom means that disentangling the data will be a complex task as discussed further below.

Assuming that the noise level for sensing gain fluctuations $\Delta R/R \sim 10\%$, the most steeply varying modes in our simulations allow radius of curvature control to about 1ppm, corresponding to about 2mm radius of curvature precision, which corresponds to optical path length control $\sim 2\times10^{-13}$m and wave front errors $\sim 2\times10^{-7}\lambda$.

The multiple channels of data that are predicted to be available in cavities with large test masses correspond to hundreds of acoustic modes which sample different test masses, their thermal distortion and their alignment. We suggest that monitoring these channels can allow improved control of high power interferometers, helping to predict and prevent parametric instability, and also possibly helping to minimize glitches that can occur due to the interaction of varying diameter beams with loss points on the mirrors.

However it is important to note that complete diagonalization (to allow extraction of individual error parameters in an advanced interferometer) would require a sufficient number of independent modes. This point is uncertain, so it is not clear how many degrees of freedom could actually be controlled.

Clearly experimental testing of these ideas is required. Because there are so many acoustic mode channels there will be a need to determine which channels to use and how to combine them to create useful tools for separate purposes such as individual test mass alignment and thermal distortion probes. Substantial software development will be required.

Experimentally the monitoring is simple, in the first instance by monitoring the transmitted light of interferometer cavities with a quadrant photo detector. Such a detector would not be sensitive to all possible transverse modes but should allow the basic concepts to be verified, and difficulties identified. The main requirement of the optical detection system has sufficient bandwidth to encompass the relevant ultrasonic modes which are up to 150kHz. Experiments could be undertaken during detector commissioning to create a database of high gain three mode interactions. Much more work combining experiment and modelling will be required to determine how to extract useful error signals.

**Acknowledgements**
The authors wish to thank Slawomir Gras for his modelling results used in this paper and useful discussions. We also thank the LIGO Scientific Collaboration (LSC), especially the LSC optical working group and Gregg Harry for the encouragement and discussions. We also thank the Gingin Advisory Committee of the LSC, chaired recently by Stefen Gossler for help and advice. This work is a project of the Australian Consortium for Interferometric Gravitational Astronomy (ACIGA) and was supported by the Australian Research Council.